\documentclass[
 reprint,
nofootinbib,
 amsmath,amssymb,
 aps,
 prd,
]{revtex4-2}

\usepackage{dcolumn}% Align table columns on decimal point
\usepackage{indentfirst}
\usepackage{amsmath}
\usepackage{bm}
\usepackage{slashed}
\usepackage{graphicx}
\usepackage{hyperref}
\usepackage{multirow}

\begin{document}

\preprint{APS/123-QED}

% \title{Mitigating cosmological tensions via coupled dark sector models}
\title{Coupled dark sector models and cosmological tensions}

\author{Gang Liu}
 \email{liugang\_dlut@mail.dlut.edu.cn}

\author{Jiaze Gao}
\author{Yufen Han}
\author{Yuhao Mu}

\author{Lixin Xu}
 \email{lxxu@dlut.edu.cn}
\affiliation{%
 Institute of Theoretical Physics\\
 School of Physics\\
 Dalian University of Technology\\
 Dalian 116024, People's Republic of China
}

\date{\today}

\begin{abstract}
In this paper, we introduce two coupling models of early dark energy (EDE) and cold 
dark matter aimed at alleviating cosmological tensions. We utilize the EDE component 
in the coupling models to relieve the Hubble tension, while leveraging the interaction 
between dark matter and dark energy to alleviate the large-scale structure tension. 
The interaction is implemented in the form of pure momentum coupling and Yukawa coupling.
We employed various cosmological datasets, including cosmic microwave background 
radiation, baryon acoustic oscillations, Type Ia supernovae, the local distance-ladder data 
(SH0ES), and the Dark Energy Survey Year-3 data, to analyze our models. 
We first \textit{exclude} SH0ES data from the entire dataset to constrain the parameters 
of novel models. We observe that the constraints on $H_0$ from two coupling models are 
slightly higher than that from the $\Lambda$CDM model, but they exhibit a significant 
inconsistency with the SH0ES data, consistent with prior research findings in the EDE model. 
Subsequently, we incorporate SH0ES data to re-constrain the parameters of various models, 
our findings reveal that both coupling models yield best-fit 
values for $H_0$ approximately around $72.23$ km/s/Mpc, effectively mitigating the Hubble tension. 
Similar to the EDE model, the coupling models yield the $S_8$ values that 
still surpasses the result of the $\Lambda$CDM model. Nevertheless, the best-fit values for 
$S_8$ obtained with the two new models are 0.8192 and 0.8177, respectively, which are lower than 
the 0.8316 achieved by the EDE model. Consequently, although our coupling models fail 
to fully resolve the large-scale structure tension, they partially mitigate the adverse 
effect of the original EDE model. 
\end{abstract}

\maketitle

%\tableofcontents

\section{Introduction}
Despite the success of the $\Lambda$CDM model in explaining various cosmological data 
such as Cosmic Microwave Background (CMB), Baryon Acoustic Oscillation (BAO), and 
Type Ia Supernovae (SNIa), it does not provide insights into the nature of dark 
matter and dark energy. Furthermore, with the increasing precision and abundance of 
cosmological observations, inconsistencies between the concordance cosmological model 
and observational data have become more pronounced. 

Among these disparities, the most renowned one is the Hubble tension, which refers to 
the inconsistency between the inferred value of the Hubble constant at high redshift 
based on CMB observations within the framework of the $\Lambda$CDM model, and the 
model-independent measured value of the Hubble constant at low redshift \cite{Verde_2019}. 

Based on the \textit{Planck} 2018 CMB data, the $\Lambda$CDM model infers the Hubble 
constant value of $67.37\pm0.54$ km/s/Mpc \cite{planck2020}. However, 
utilizing the distance ladder method based on cepheid-calibrated SNIa data, the SH0ES 
measurement yields the Hubble constant value of $73.04\pm1.04$ km/s/Mpc 
\cite{Riess_2022}, resulting in a statistical error of 4.8$\sigma$.

Another manifestation of a relatively mild tension concerns the contradiction between 
measurements of large-scale structure and CMB \cite{PRL.111.161301, Hildebrandt_2020}, 
typically described by $S_8\equiv\sigma_8\sqrt{(\Omega_\mathrm{m}/0.3)}$. Here, $\Omega_m$ 
represents the current total matter energy density fraction, while $\sigma_8$ denotes 
the root mean square of matter fluctuations at a scale of 8 $h^{-1}$Mpc. The \textit{Planck} 
2018 best-fit $\Lambda$CDM model predicts the $S_8$ value of $0.834\pm0.016$ \cite{planck2020}. 
However, measurements from large-scale structure, such as the Dark Energy Survey 
Year-3 (DES-Y3), yields the $S_8$ value of $0.776\pm0.017$ \cite{PRD.105.023520}.

Various models have been proposed to address the issues concerning dark matter and dark 
energy. Commonly encountered models include various dynamical dark energy models 
\cite{Guo_2019, Li_2019, Zhouzh}, early dark energy  \cite{PhysRevLett.122.221301,
PRD.102.043507,PhysRevD.101.063523}, new early dark energy \cite{PhysRevD.103.L041303,
PhysRevD.102.063527}, decaying dark matter \cite{Buch_2017, PRD.98.023543, PRD.103.043014, 
Alvi_2022}, interacting dark matter \cite{PRD.105.103509}, axion dark matter 
\cite{D'Eramo_2018,liu2023cosmological}, interacting dark energy 
\cite{sty2789, Di_Valentino_2020} and so on.

One of the most intriguing models is the early dark energy (EDE) model 
\cite{PhysRevLett.122.221301,PRD.102.043507,PhysRevD.101.063523}. By introducing 
an EDE component before recombination, it is possible to reduce the comoving sound 
horizon of last scattering, 
\begin{equation}
    r_s(z_*)=\int_{z_*}^{\infty}\frac{c_s(z)}{H(z)}dz, 
\end{equation}
where $z_*$ represents the redshift of last scattering, and $c_s$ is the speed of 
sound of the photon-baryon plasma. This allows for an increase in $H_0$ while maintaining 
consistency with the CMB observations of angular scale of the sound horizon, 
\begin{equation}
    \theta_s=\frac{r_s(z_*)}{D_A(z_*)}, 
\end{equation}
where, $D_A(z_*)$ refers to the angular diameter distance to the last scattering. 

\begin{figure}
	\includegraphics[width=\columnwidth]{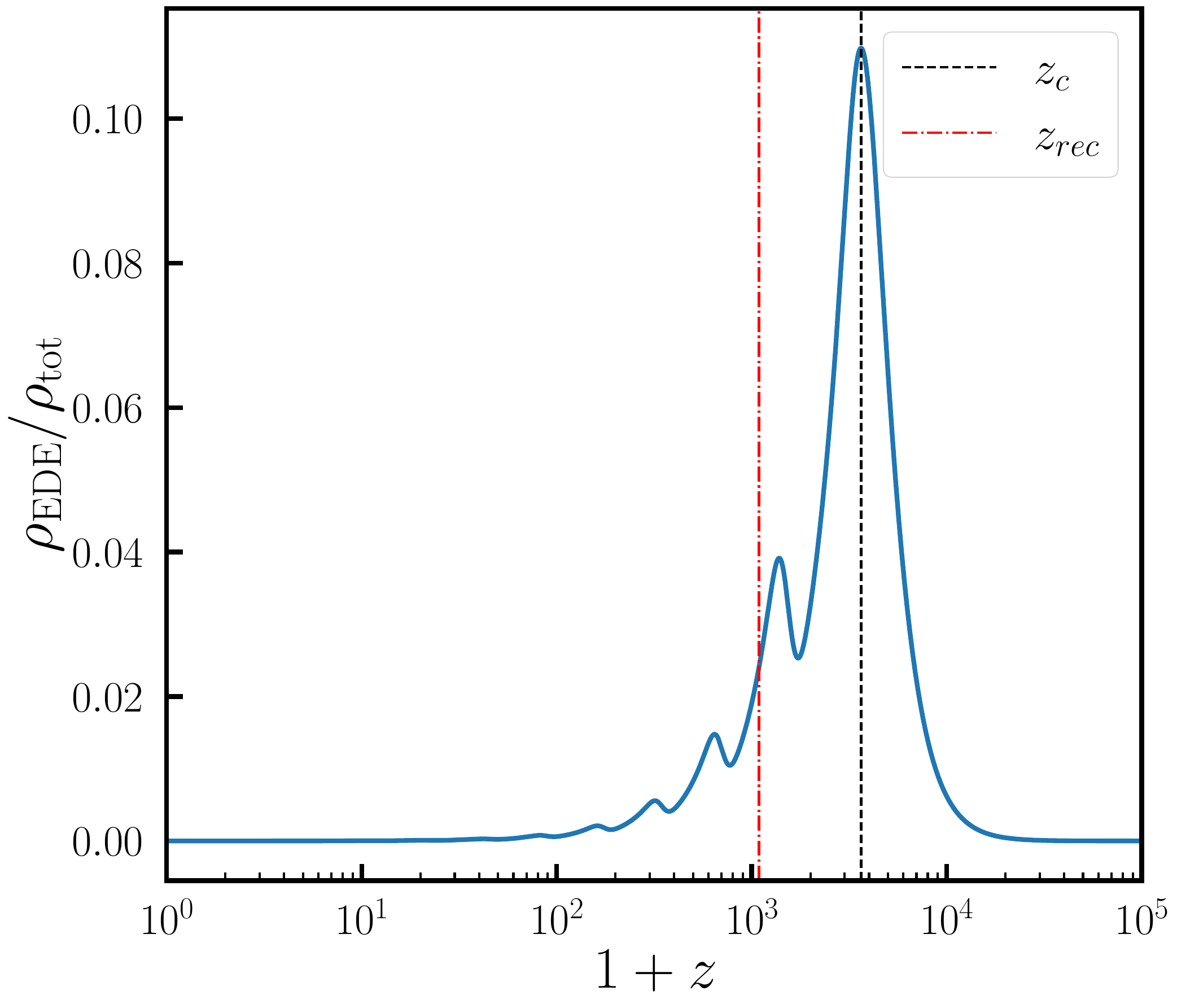}
    \caption{The evolution of the energy density fraction of EDE with respect to redshift. 
        The red dash-dotted line represents the recombination redshift, and it can be observed 
        that the peak contribution of EDE occurs before recombination.}
    \label{fig:1}
\end{figure}

EDE is typically described by an ultra-light axion scalar field \cite{PhysRevLett.113.251302,
MARSH20161}. We denote the redshift corresponding to the peak contribution of 
EDE as $z_c$ and the ratio of EDE energy density to the total energy density at 
this redshift as $f_\mathrm{EDE}$. Figure~\ref{fig:1} illustrates the evolution of 
the EDE energy density fraction with redshift, where the red dash-dotted line 
represents the recombination redshift and the EDE contribution peak occurs 
earlier than recombination. The parameters used in the figure are shown in 
Eq.~(\ref{eq:ede}). 
% The Hubble tension can be resolved when 
% $f_\mathrm{EDE}$ reaches approximately 10\%. 

Despite the partially mitigation of the Hubble tension by the EDE model, it introduces 
additional issues. The EDE component suppresses the growth of perturbations during its 
contribution period, necessitating an increase in the cold dark matter density to remain 
consistent with CMB data. Furthermore, several other cosmological parameters, such as the 
scalar spectral index $n_\mathrm{s}$, baryon density $\omega_\mathrm{b}$, and amplitude of density 
fluctuations $\sigma_8$, undergo changes \cite{PhysRevD.106.043525}. As a 
result, the EDE model further exacerbates the existing large-scale structure 
tension \cite{PRD.102.043507,D'Amico_2021}.

Our primary focus is on the EDE model whereby the interaction between 
dark matter and EDE is introduced to mitigate the adverse effect associated with the 
EDE model. Previous works have investigated various forms of interactions between dark 
matter and EDE \cite{PhysRevD.106.043525,PhysRevD.107.103523,liu2023alleviating, 
liu2023kinetically}. In this paper, we discuss two forms of coupling, pure momentum 
coupling and Yukawa coupling. 

The pure momentum interaction between dark matter and dark energy has been investigated 
in some previous coupled quintessence models \cite{PhysRevD.88.083505,PhysRevD.94.043518,
PhysRevD.101.043531}. In contrast to many previous phenomenological interacting dark 
energy models \cite{Caldera-Cabral_2009,Koyama_2009,Benetti_2019,PhysRevD.104.103503}, 
the authors in \cite{PhysRevD.88.083505} utilize the pull-back formalism to provide a 
generalized fluid action that includes scalar field couplings, where the \textit{Type 3} model 
correspond to theory of pure momentum transfer.
In this investigation, we extend its application to the coupling between early dark 
energy and cold dark matter, proposing the momentum-coupled dark sector (MCDS) model. 

The Yukawa coupling form was originally utilized to describe the 
interaction between pions and nucleons \cite{10.1143/PTPS.1.1}. Here, we extend its application to 
characterize the interaction between dark matter and dark energy \cite{Farrar_2004,Archidiacono_2022}, 
where the scalar field represents the dark energy component, while the fermion field represents 
the dark matter component. We employ this form of interaction to construct the Yukawa-coupled dark 
sector (YCDS) model.

The EDE component in the coupling models is introduced to alleviate the Hubble 
tension, while the interaction between cold dark matter and EDE is employed to 
suppress the growth of matter structures and thereby alleviate large-scale structure 
tension. 

The structure of this paper is organised as follows: In Sec.~\ref{sec:mo}, we present 
two coupling models and provide the background and perturbation evolution equations for EDE 
and cold dark matter. Section~\ref{sec:nr} presents the numerical results, including the 
impact on the Hubble parameter and matter power spectrum. In Sec.~\ref{sec:dm}, we 
introduce the datasets used for the Monte Carlo Markov Chain analysis and present the 
constrained outcomes. Finally, in Sec.~\ref{sec:con}, we summarize our findings. 

\section{Two coupling dark sector models}
\label{sec:mo}
The action of early dark energy (EDE), cold dark matter, and interaction term can be 
represented as follows,
\begin{footnotesize}
\begin{equation}
    \mathcal{S}=\int d^4x\sqrt{-g}\left[
    -\frac{1}{2}\partial^{\mu}\phi\partial_{\mu}\phi-V(\phi)-i\bar{\psi}\slashed{D}\psi
    -m_{\psi}\bar{\psi}\psi+\mathcal{L}_\mathrm{int}\right],
    \label{eq:act}
\end{equation}
\end{footnotesize}
where $\phi$ represents the EDE scalar and $\psi$ is the Dirac fermion that 
plays the role of cold dark matter, with $m_\psi$ denoting its mass. 
In the non-relativistic limit, 
$\langle \bar{\psi}\psi \rangle \to n_{\psi}$, where $n_{\psi}$ represents the number 
density. The energy-momentum tensor of cold dark matter can be expressed as, 
\begin{equation}
    T^{\mu}_{(c)\nu}=m_{\psi} n_{\psi} u^{\mu}u_{\nu}=\rho_c u^{\mu}u_{\nu},
\end{equation}
where $u^{\mu}$ and $\rho_c$ denote the four-velocity and energy density of cold dark 
matter, respectively. We employ the subscript ``$_c$'' to symbolize cold dark matter 
in the subsequent discourse. 
We adopt the EDE potential form from \cite{PRD.102.043507, PhysRevD.101.063523}, 
\begin{equation}
    V(\phi)=m_{\phi}^2f_{\phi}^2[1-\cos(\phi/f_{\phi})]^3+V_{\Lambda},
\end{equation}
where $m_{\phi}$ denotes the axion mass, $f_{\phi}$ represents the decay constant, and 
$V_{\Lambda}$ performs as the cosmological constant.

\subsection{Momentum-coupled dark sector model}
We have the flexibility to select the form of interactions in order to obtain various 
specific models. Following \cite{PhysRevD.88.083505,PhysRevD.94.043518}, we firstly focus on 
the following pure momentum coupling form, 
\begin{equation}
    \mathcal{L}_\mathrm{int}=-\beta(u^{\mu}\partial_{\mu}\phi)^2,
\end{equation}
where $\beta$ is a constant that describes the strength of the coupling. Consequently, 
the EDE and cold dark matter solely engage in momentum exchange. 

By varying the action Eq.~(\ref{eq:act}) with respect to the metric $g^{\mu\nu}$, 
we obtain the energy-momentum tensor for EDE and cold dark matter, including their 
interaction term, as follows,
\begin{equation}
    \begin{aligned}
        T^{\mu}_{(\phi)\nu}+T^{\mu}_{(c)\nu}&=
        \partial^{\mu}\phi\partial_{\nu}\phi+\rho_c u^{\mu}u_{\nu}
        +2\beta(u^{\alpha}\partial_{\alpha}\phi)(u^{\mu}\partial_{\nu}\phi)\\
        &-\delta^{\mu}{}_{\nu}[\frac{1}{2}\partial^{\alpha}\phi\partial_{\alpha}\phi+V(\phi)+\beta(u^{\alpha}\partial_{\alpha}\phi)^2],
    \end{aligned}
    \label{eq:Tmn}
\end{equation}
with $u^{\mu}=a^{-1}(1,v^i),\, u_{\mu}=a(-1,v^i)$, where $a$ denotes the scale factor and 
$v^i$ is the three-velocity of 
cold dark matter. Due to the consideration of only the interaction between EDE and cold 
dark matter, the total energy-momentum tensor of both components is covariantly conserved, 
\begin{equation}
    \nabla_{\mu}\left(T^{\mu}_{(\phi)\nu}+T^{\mu}_{(c)\nu}\right)=0.
    \label{eq:bian}
\end{equation}

We decompose the EDE scalar and the energy density of cold dark matter into their 
background and perturbation components, 
\begin{subequations}
    \begin{align}
        &\phi=\bar{\phi}(\tau)+\delta\phi(\tau,x^i),\\
        &\rho_c=\bar{\rho}_c+\delta \rho_c=\bar{\rho}_c(1+\delta_c),
    \end{align} 
\end{subequations}
where $\tau$ represents conformal time. 

\subsubsection{Background equations}
The variation of the action expanded to linear order in $\delta\phi$ yields the equation 
of motion for the scalar field background, 
\begin{equation}
    \bar{\phi}''+2\mathcal{H}\bar{\phi}'+\frac{a^2V_{\phi}}{1-2\beta}=0,
    \label{eq:kg}
\end{equation}
where the prime denotes the derivative with respect to conformal time $\tau$, $\mathcal{H}$ 
is the conformal Hubble parameter, and $V_{\phi}$ denotes the partial derivative of the 
EDE potential with respect to $\bar{\phi}$.

By evaluating Eq.~(\ref{eq:bian}) at the background level and substituting the 
result of Eq.~(\ref{eq:kg}) into it, we obtain the energy density equation for cold 
dark matter,
\begin{equation}
    \bar{\rho}_c'=-3\mathcal{H}\bar{\rho}_c.
\end{equation}

If we define the energy density and pressure of momentum-coupled EDE in the following form,
\begin{subequations}
    \begin{align}
        &\bar{\rho}_{\phi}=\frac{\bar{\phi}'^2}{2a^2}(1-2\beta)+V(\bar{\phi}),\\
        &\bar{p}_{\phi}=\frac{\bar{\phi}'^2}{2a^2}(1-2\beta)-V(\bar{\phi}),
    \end{align} 
    \label{eq:rhop}
\end{subequations}
the energy density equation for EDE is given by, 
\begin{equation}
    \bar{\rho}_{\phi}'=-3\mathcal{H}(\bar{\rho}_{\phi}+\bar{p}_{\phi}).
\end{equation}

We can observe that the continuity equations for EDE and cold dark matter are consistent 
with their uncoupled forms. This is because we only consider the momentum exchange 
between EDE and cold dark matter, which only affects their velocity evolution equations. 

In addition, by combining Eq.~(\ref{eq:kg}) and Eq.~(\ref{eq:rhop}), it can be 
inferred that the model is physically viable for $\beta<\frac{1}{2}$. When 
$\beta\to\frac{1}{2}$, a strong coupling issue arises. For $\beta>\frac{1}{2}$, 
the presence of a negative kinetic term leads to the inclusion of ghost in the 
theory \cite{PhysRevD.94.043518}.

\subsubsection{Perturbation equations}
We utilize the synchronous gauge to derive the perturbation equations for EDE and cold 
dark matter, the line element is defined as,
\begin{equation}
    ds^2=a^2(\tau)\left[-d\tau^2+(\delta_{ij}+h_{ij})dx^idx^j\right].
\end{equation}
The variation of the action expanded to quadratic order in $\delta\phi$ yields the 
equation of motion for the scalar field perturbation,
\begin{equation}
    \delta\phi''+2\mathcal{H}\delta\phi'+\frac{1}{2}h'\bar{\phi}'+\frac{(k^2+a^2V_{\phi\phi})\delta\phi}{1-2\beta}-\frac{2\beta\bar{\phi}'\theta_c}{1-2\beta}=0,
\end{equation}
where $V_{\phi\phi}$ represents the second-order partial derivative of the EDE 
potential with respect to $\bar{\phi}$, and $\theta_c\equiv\partial_i v^i$ is the 
velocity divergence of cold dark matter.

According to Eq.~(\ref{eq:Tmn}), the density perturbation, 
pressure perturbation, and velocity divergence of momentum-coupled EDE are given by,
\begin{subequations}
    \begin{align}
        &\delta\rho_{\phi}=\frac{\bar{\phi}'\delta\phi'}{a^2}(1-2\beta)+V_{\phi}\delta\phi,\\
        &\delta p_{\phi}=\frac{\bar{\phi}'\delta\phi'}{a^2}(1-2\beta)-V_{\phi}\delta\phi,\\
        &(\bar{\rho}_{\phi}+\bar{p}_{\phi})\theta_{\phi}=\frac{\bar{\phi}'}{a^2}k^2\delta\phi(1-2\beta).
    \end{align} 
\end{subequations}

By calculating Eq.~(\ref{eq:bian}) at the perturbation level, we obtain the density 
contrast and velocity evolution equations for cold dark matter, 
\begin{subequations}
    \begin{align}
        &\delta_c'+\theta_c+\frac{1}{2}h'=0,\\
        &\theta_c'+\mathcal{H}\theta_c=\frac{2\beta\mathcal{H}\bar{\phi}'}{a^2\bar{\rho}_c}\left(\bar{\phi}'\theta_c-k^2\delta\phi\right).
    \end{align} 
\end{subequations}
It can be observed that the equation for the density contrast of cold dark matter is 
consistent with its uncoupled form, while the velocity equation is coupled with EDE.
The coupled model indirectly affects the density contrast equation of cold dark matter 
by modifying its velocity equation, thereby suppressing the growth of structures and 
alleviating large-scale structure tension.

\subsection{Yukawa-coupled dark sector model}
We can also adopt the Yukawa interaction form to describe the coupling between EDE and cold dark matter,
\begin{equation}
    \mathcal{L}_\mathrm{int}=-\kappa\phi\bar{\psi}\psi,
\end{equation}
where $\kappa$ represents the dimensionless Yukawa coupling constant that describes the strength of the 
interaction. We can absorb the interaction term into the potential term of the fermion field. Specifically, 
if we use the following transformation form, 
\begin{equation}
    \kappa=\xi\frac{m_\psi}{M_{pl}},
\end{equation}
where $\xi$ is a dimensionless constant and $M_{pl}$ represents the reduced Planck mass, 
then the mass of cold dark matter including the coupling term can be expressed as,
\begin{equation}
    m_c=m_\psi+\kappa\phi=m_\psi(1+\frac{\xi\phi}{M_{pl}}).
\end{equation}

The energy density of cold dark matter is given by, 
\begin{equation}
    \rho_c=m_c n_{\psi}=\widetilde{\rho_c}(1+\frac{\xi\phi}{M_{pl}}),
\end{equation}
where $\widetilde{\rho_c}$ represents the energy density of cold dark matter without interaction. 
We find that the Yukawa coupling of the dark sector is equivalent to the dependence of the cold dark matter 
energy density on the EDE scalar. Previous research has utilized the swampland conjecture 
\cite{Vafa2005TheSL,Palti} to propose a coupling form where the dark matter energy density exhibits 
exponential dependence on the EDE scalar \cite{PhysRevD.106.043525,liu2023alleviating}. The Yukawa 
coupling model can be regarded as a higher-order truncation of the exponential form.

\subsubsection{Background Equations}
Expanding the action in Eq.(\ref{eq:act}) to linear order and carrying out the 
variation with respect to $\delta\phi$, we obtain the background evolution equation for the EDE scalar,
\begin{equation}
    \bar{\phi}''+2\mathcal{H}\bar{\phi}'+a^2V_{\phi}=-a^2F\bar{\rho}_c,
    \label{eq:kgb}
\end{equation}
where 
\begin{equation}
    F=\frac{\xi}{M_{pl}+\xi\bar{\phi}}.
\end{equation}

By utilizing the conservation of the total energy-momentum tensor for dark matter and dark energy, in conjunction 
with Eq.(\ref{eq:bian}) and Eq.(\ref{eq:kgb}), we obtain the energy density equation for cold dark matter,
\begin{equation}
    \bar{\rho}'_c+3\mathcal{H}\bar{\rho}_c=\bar{\phi}'F\bar{\rho}_c.
\end{equation}

The energy density and pressure of EDE are defined as follows \cite{PRD.58.023503},
\begin{subequations}
    \begin{align}
        &\bar{\rho}_{\phi}=\frac{\bar{\phi}'^2}{2a^2}+V(\bar{\phi}),\\
        &\bar{p}_{\phi}=\frac{\bar{\phi}'^2}{2a^2}-V(\bar{\phi}).
    \end{align} 
    \label{eq:rhop2}
\end{subequations}
Combining the Klein-Gordon equation for the EDE scalar field in Eq.(\ref{eq:kgb}), we obtain the energy density 
evolution equation for EDE, 
\begin{equation}
    \bar{\rho}_{\phi}'+3\mathcal{H}(\bar{\rho}_{\phi}+\bar{p}_{\phi})=-\bar{\phi}'F\bar{\rho}_c.
    \label{eq:rb}
\end{equation}

\begin{figure*}
    \includegraphics[width=\linewidth]{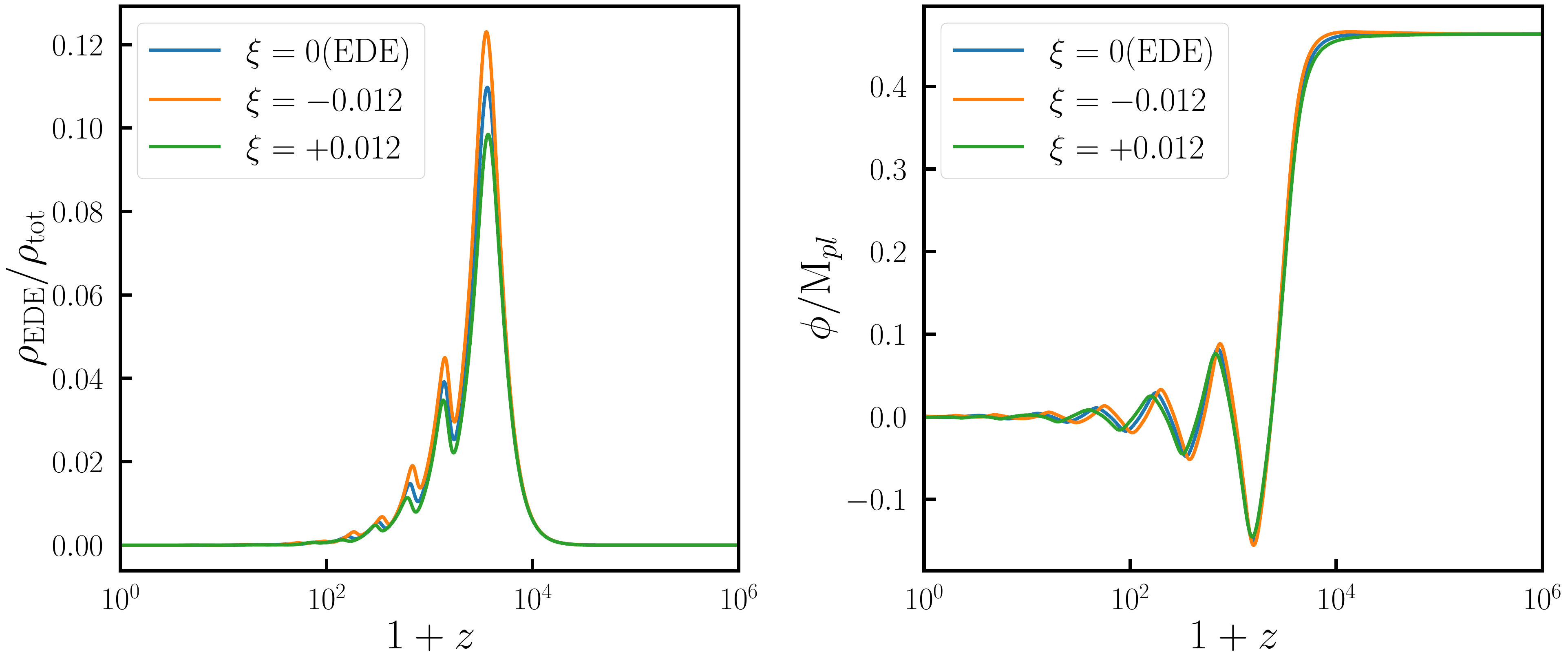}
    \caption{The evolution with redshift of the EDE energy density fraction (left panel) and the EDE scalar (right 
    panel) for various coupling constants is presented. The EDE energy density fraction and the amplitude and 
    phase of the EDE scalar are affected by different coupling constants. A negative coupling constant leads to 
    an increase in the EDE energy density fraction.}
\label{fig:2}
\end{figure*}

In Fig.~\ref{fig:2}, we illustrate the evolution with redshift of the ratio of EDE energy density to total 
energy density (left panel) and the EDE scalar (right panel) for different coupling constants. The remaining 
cosmological parameters are taken from Eq.~(\ref{eq:ede}). Different coupling constants lead to variations in 
the EDE energy density fraction and the amplitude and phase of the EDE scalar. The sign of the coupling 
constant determines the direction of energy density transfer between dark matter and dark energy. A negative 
coupling constant results in energy transfer from dark matter to dark energy, leading to a greater EDE energy 
density fraction, while the effect is reversed for a positive coupling constant.

\subsubsection{Perturbation equations}
Expanding the action to the quadratic order and taking the variation with respect to $\delta\phi$, we can derive the 
perturbation evolution equation for the EDE scalar,
\begin{small}
\begin{equation}
    \delta\phi''+2\mathcal{H}\delta\phi'+\frac{1}{2}h'\bar{\phi}'+(k^2+a^2V_{\phi\phi})\delta\phi
    =-a^2F\bar{\rho}_c(\delta_c-F\delta\phi).
\end{equation}
\end{small}

By exploiting the covariant conservation of the total energy-momentum tensor for cold dark matter and dark energy, 
we can derive the evolution equations for the density contrast and velocity divergence of cold dark matter,
\begin{subequations}
    \begin{align}
        &\delta_c'+\theta_c+\frac{1}{2}h'=F(\delta\phi'-F\bar{\phi}'\delta\phi),\\
        &\theta_c'+\mathcal{H}\theta_c=F(k^2\delta\phi-\bar{\phi}'\theta_c).
    \end{align} 
\end{subequations}

\subsection{Initial conditions}
In the early universe, the Hubble friction in the scalar field dominates, leading to 
effective freezing of the EDE scalar, with the initial value of $\bar{\phi}'$ set to 0. We 
take the ratio of the initial values of $\bar{\phi}$ and the axion decay constant, 
$\alpha_i\equiv\bar{\phi}_i/f_{\phi}$, as the model parameter \cite{PRD.102.043507,PhysRevD.101.063523}. 
As for cold dark matter, the background evolution equations for the energy density of cold dark matter degenerate to the 
form of the non-coupled case. Hence, we do not alter the initial conditions for cold dark matter. When calculating the 
perturbation equations, we employ adiabatic initial conditions, keeping the initial 
conditions for cold dark matter unchanged, and referring to \cite{PhysRevD.101.063523} 
for the initial conditions of EDE.

\section{Numerical results}
\label{sec:nr}
Based on the description in the previous section, we made modifications to the 
publicly available Boltzmann code \texttt{CLASS} 
\cite{Blas_2011,1104.2932}. We have incorporated a new component of cold dark matter into the 
calculation of the velocity equation, accounting for the coupling effects. Furthermore, we 
retained the original cold dark matter 
component within \texttt{CLASS} and set $\Omega_\mathrm{cdm}=10^{-6}$ to maintain 
consistency with the definition of the synchronous gauge.

We present numerical results with cosmological parameters adopted from Table IV in 
\cite{PhysRevD.106.043525}. Specifically, for the $\Lambda$CDM model, we employ the 
following parameter values:
\begin{align}
    &100\theta{}_\mathrm{s}=1.04202, \quad \omega_\mathrm{b}=0.02258,\\
    \notag
    &\omega_\mathrm{c}=0.1176, \quad \ln(10^{10}A_\mathrm{s})=3.041,\\
    \notag
    &n_\mathrm{s}=0.9706, \quad \tau_\mathrm{reio}=0.0535.
\end{align}
For the two coupling models, we exclusively vary the values of the coupling parameters, while 
keeping the values of other cosmological parameters fixed to the constraints obtained 
from the EDE model,
\begin{align}
    \label{eq:ede}
    &100\theta{}_\mathrm{s}=1.04138, \quad \omega_\mathrm{b}=0.02281,\\
    \notag
    &\omega_\mathrm{c}=0.1287, \quad \ln(10^{10}A_\mathrm{s})=3.065,\\
    \notag
    &n_\mathrm{s}=0.9895, \quad \tau_\mathrm{reio}=0.0581, \quad \alpha_i=2.77,\\
    \notag
    &\log_{10}(f_{\phi})=26.61, \quad \log_{10}(m_{\phi})=-27.31.    
\end{align}

\subsection{Momentum-coupled dark sector model}

\begin{figure}
	\includegraphics[width=\columnwidth]{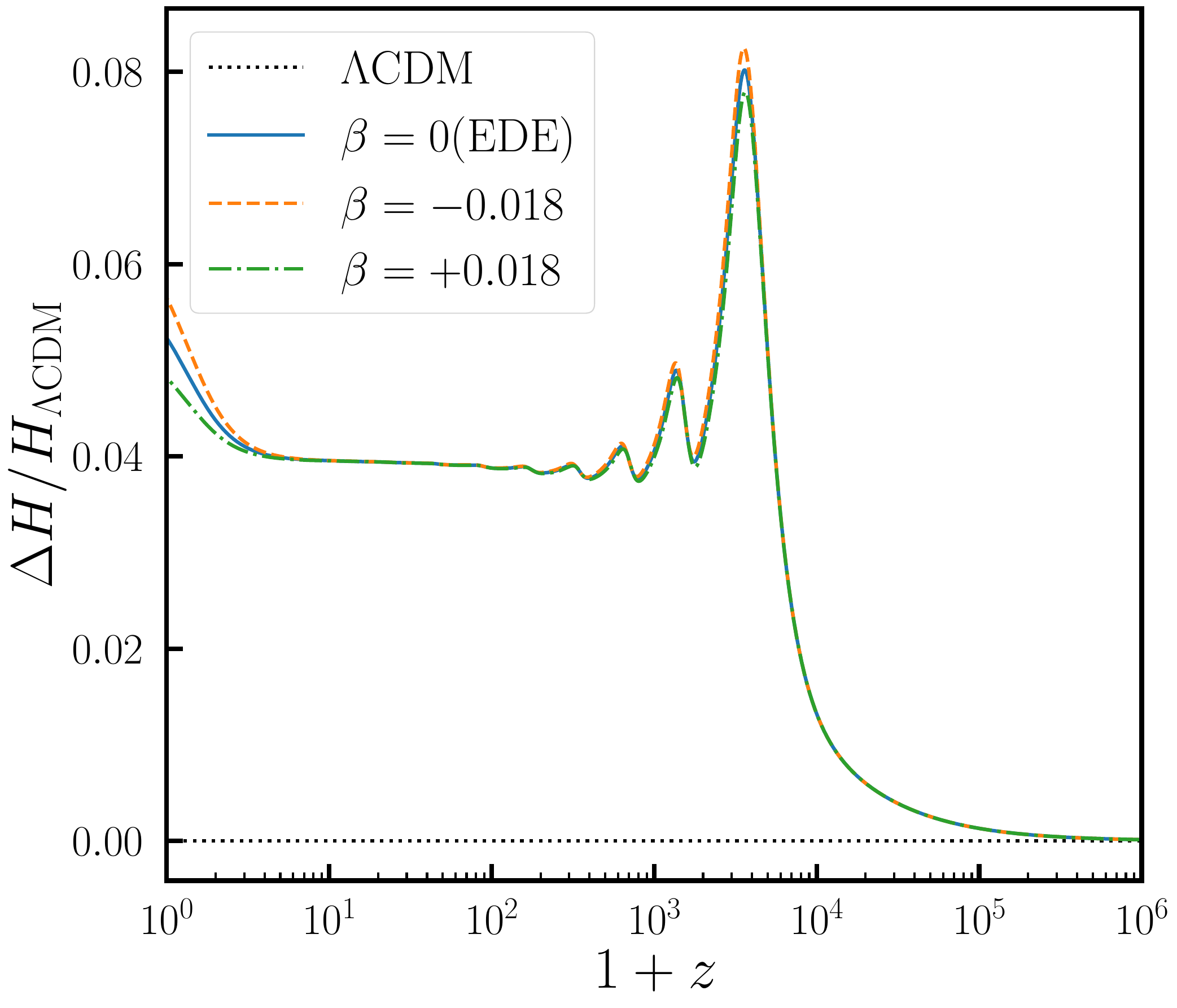}
    \caption{The evolution of the Hubble parameter with redshift. The EDE component in the 
        MCDS model enhances the Hubble parameter. A negative coupling parameter amplifies 
        the effect of EDE, while positive coupling parameters weaken it.}
    \label{fig:3}
\end{figure}

We demonstrate in Fig.~\ref{fig:3} the impact of different values of the coupling parameter 
$\beta$ in the MCDS model on the evolution of the Hubble parameter.
The black dotted line represents the $\Lambda$CDM model, while the blue solid, orange 
dashed, and green dash-dotted lines represent the results for the MCDS model with 
coupling parameters 0, $-0.018$, and 0.018, respectively. It is worth noting that the MCDS 
model degenerates to the EDE model when the coupling parameter is set to 0. 

It is evident that the EDE component in the MCDS model increases the Hubble parameter. 
Furthermore, a negative coupling parameter further enhances the Hubble parameter relative 
to the EDE model, while positive coupling parameters have the opposite effect. This 
phenomenon can be easily explained by the energy density formula for EDE in 
Eq.~(\ref{eq:rhop}), where a negative coupling constant effectively amplifies the 
kinetic energy of EDE, resulting in an increased Hubble parameter. Conversely, positive 
coupling parameters diminish the Hubble parameter.

\begin{figure}
	\includegraphics[width=\columnwidth]{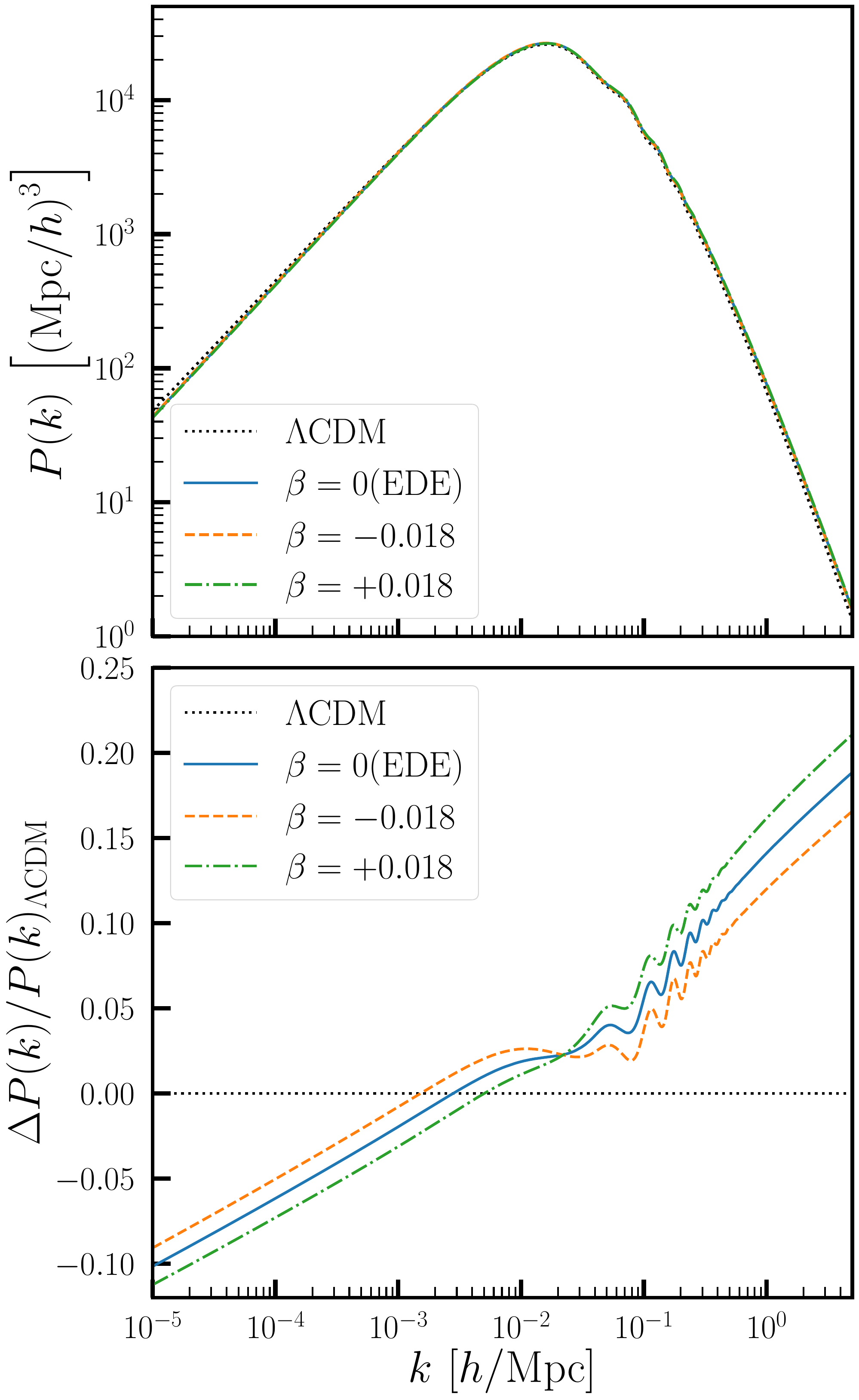}
    \caption{The linear matter power spectra (upper panel) of the MCDS model with different 
        coupling parameters, as well as the differences in power spectra relative to the 
        $\Lambda$CDM model (lower panel), are depicted. Comparing with the EDE model (with 
        the coupling constant of 0), a negative (positive) coupling constant reduces (increases) 
        the power spectrum on small scales.}
    \label{fig:4}
\end{figure}

In Fig.~\ref{fig:4}, we showcase the linear matter power spectra of the MCDS model with 
different coupling parameters (upper panel) as well as the differences in power spectra 
relative to the $\Lambda$CDM model (lower panel). 

It can be observed that the matter power spectrum of the momentum-coupled model still exceeds 
that of the $\Lambda$CDM model on small scales. 
However, the non-zero coupling constants result in momentum 
exchange between EDE and cold dark matter, which affects the velocity 
evolution equation of cold dark matter and indirectly impacts structure growth. 
Specifically, in comparison to the result of the EDE model (with the coupling constant of 0), 
a negative coupling constant can decrease the power spectrum on small scales, thereby 
mitigating the negative effect in the original EDE model. 

\subsection{Yukawa-coupled dark sector model}

\begin{figure}
	\includegraphics[width=\columnwidth]{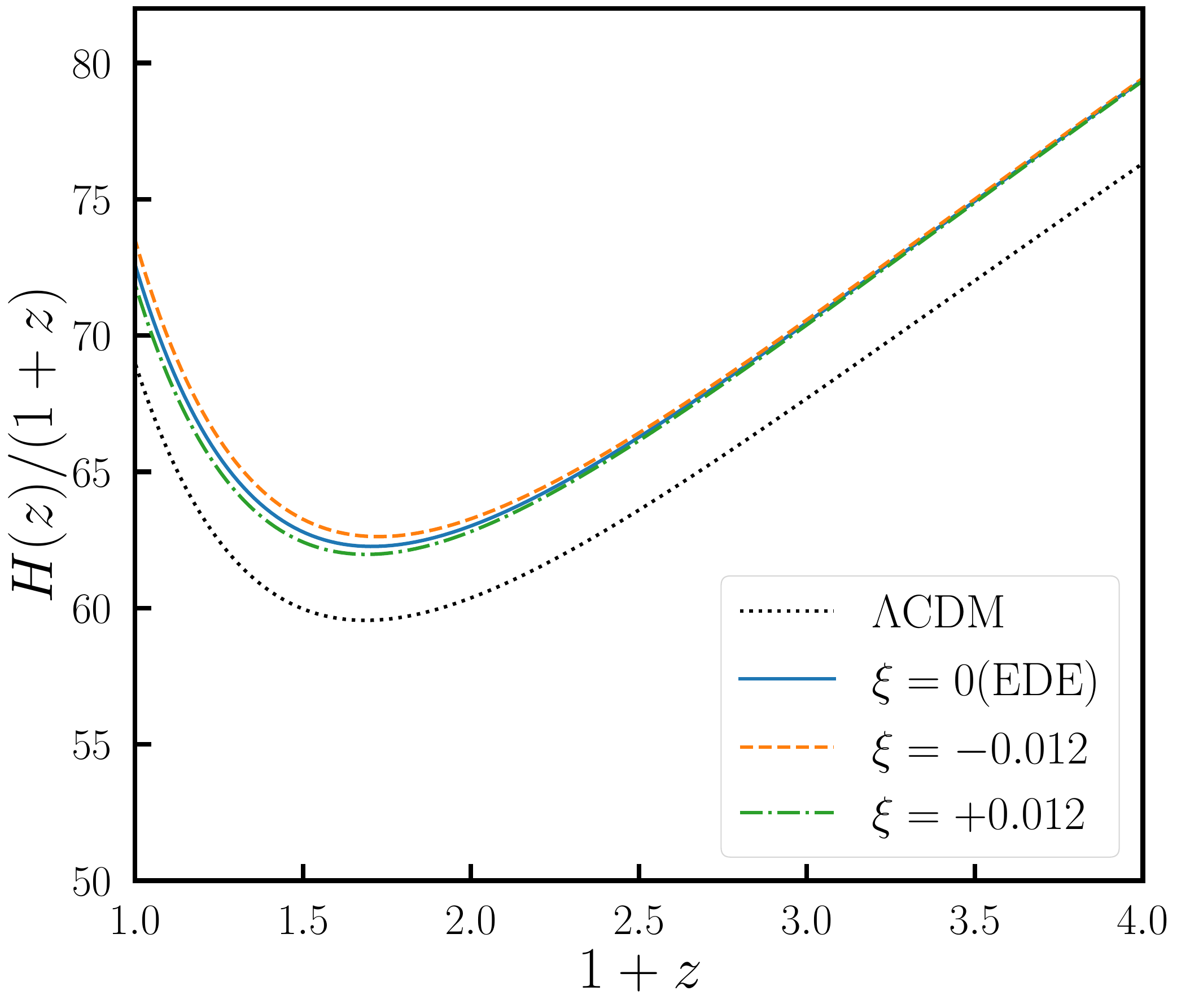}
    \caption{The Hubble parameter evolves with redshift in the YCDS model. Different coupling 
    constants impact the magnitude of the Hubble parameter based on the EDE model (with coupling 
    constant $\xi=0$). A negative value of the coupling constant increases the fraction of EDE energy 
    density, thereby indirectly increasing $H_0$. }
    \label{fig:5}
\end{figure}

Figure~\ref{fig:5} presents the redshift evolution of the Hubble parameter for the YCDS model. 
The black dotted line represents the $\Lambda$CDM model, while the blue solid line, orange dashed 
line, and green dash-dotted line correspond to the YCDS model with coupling constants $\xi$ set to 0, 
$-0.12$, and 0.12, respectively. 

The inclusion of the EDE component in the YCDS model leads to a higher Hubble 
parameter compared to the $\Lambda$CDM model. The magnitude of the Hubble parameter is further 
influenced by different coupling constants based on the EDE model (with coupling constant $\xi=0$). 
A negative coupling constant introduces source term in Eq.~(\ref{eq:rb}) for the evolution of the 
energy density of dark energy, leading to an increase in the EDE energy density fraction $f_\mathrm{EDE}$, 
(as demonstrated by the influence of 
different coupling constants on the EDE energy density fraction illustrated in Fig.~\ref{fig:2}), 
thereby indirectly leading to an augmentation in the value of $H_0$. 
Conversely, positive value of the coupling constant yield the opposite effect. 

\begin{figure}
	\includegraphics[width=\columnwidth]{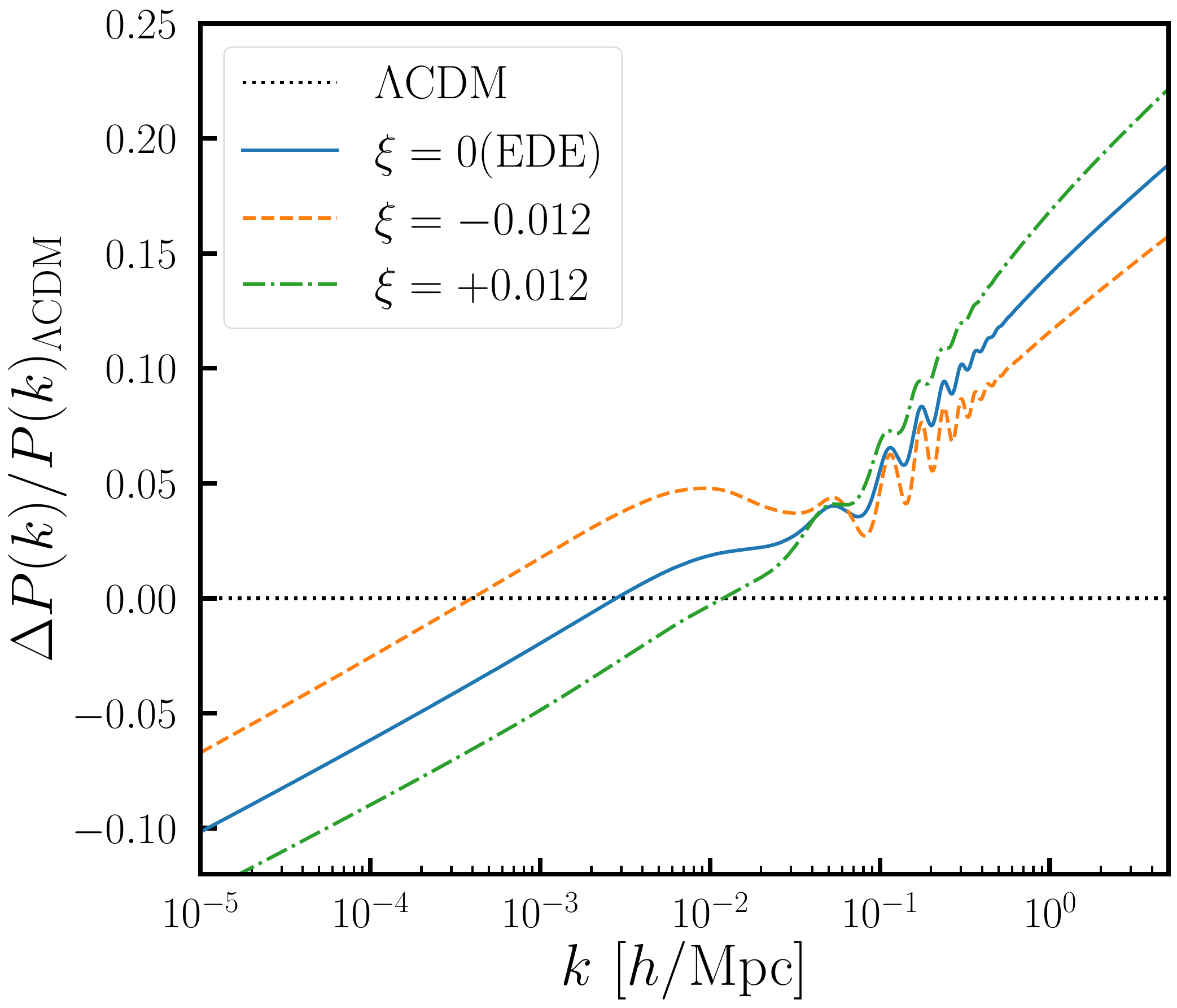}
    \caption{The differences in the linear power spectrum relative to the $\Lambda$CDM model that arise 
    from varying coupling constants in the YCDS model. A negative value of the coupling constant 
    exhibits the reduction in the matter power spectrum on small scales, while a positive value of the coupling constant 
    has the opposite effect.}
    \label{fig:6}
\end{figure}

In Fig.~\ref{fig:6}, we showcase the differences in power spectrum relative to the $\Lambda$CDM model 
when different coupling constants are taken in the YCDS model. The interaction between dark matter and 
dark energy affects the growth of matter structures and alters the shape of the power spectrum. 

A negative value of the coupling constant corresponds to the transfer of 
energy density from dark matter to dark energy, reducing the amount of dark matter, together with the drag effect 
of dark energy on dark matter, this suppresses the clustering of matter and results in smaller $P(k)$ spectra on 
small scales compared to the original EDE model. 

\section{Data and Methodology}
\label{sec:dm}
We employ \texttt{MontePython} 
\cite{Audren_2013,brinck} to perform the Markov Chain Monte Carlo (MCMC) computations 
in order to obtain the posterior distribution of model parameters. The MCMC chains are 
analyzed using \texttt{GetDist} \cite{lewis2019getdist}.

\subsection{Datasets}
In our MCMC analysis, we consider the following datasets:
\begin{itemize}
    \item[1.] \textbf{CMB}: The temperature and polarization power spectra derived from 
    the low-$\ell$ and high-$\ell$ measurements of the \textit{Planck} 2018 data, along 
    with the power spectrum of CMB lensing 
    \cite{osti_1676388,osti_1775409,planck2020}.
    \item[2.] \textbf{BAO}: The measurements acquired from the BOSS-DR12 $f\sigma_8$ 
    sample encompass the combined LOWZ and CMASS galaxy samples \citep{Alam_2017,Buen_Abad_2018}, 
    along with the low-redshift measurements derived from 6dFGS and the SDSS DR7 
    \citep{19250.x,stv154}. 
    \item[3.] \textbf{Supernovae}: The Pantheon dataset comprises 1048 type Ia supernovae 
    with redshift values spanning from 0.01 to 2.3 \citep{Scolnic_2018}.
\end{itemize} 

Through the amalgamation of CMB and BAO data, multiple acoustic horizon measurements can 
be made at different redshifts, thereby alleviating geometric degeneracies and limiting 
the physical processes between recombination and the BAO measurement redshift. Furthermore, 
the supernova data obtained from the Pantheon sample exerts a substantial constraint on 
new physics specific to the late epoch within the redshift range that is measured. 

\begin{itemize}
    \item[4.] \textbf{SH0ES}: The most recent SH0ES measurement has estimated the value 
    of the Hubble constant as $73.04 \pm 1.04$ km/s/Mpc \cite{Riess_2022}.
\end{itemize}

We utilize the $H_0$ measurements obtained from SH0ES to mitigate the influence of the 
prior volume effect \cite{PRD.103.123542} and evaluate the effectiveness of the novel 
model in addressing the tension between the local measurement of $H_0$ and the inference 
results from CMB analysis. 

\begin{itemize}
    \item[5.] \textbf{DES-Y3}: Dark Energy Survey Year-3 weak lensing and galaxy cluster 
    data, with a Gaussian constraint on $S_8$ of $0.776 \pm 0.017$ \cite{PRD.105.023520}.
\end{itemize}

We incorporate the $S_8$ data from DES-Y3 to 
investigate how well the model performs in alleviating the large-scale structure tension. 
Previous studies have validated the effectiveness of using the $S_8$ prior approach to 
approximate DES-Y1 data in the context of EDE \cite{PRD.102.043507}. In this study, we 
assume that the $S_8$ prior remains a good approximation when using DES-Y3 data for the 
EDE model. Additionally, we propose two coupling dark sector models that exhibit only minor 
deviations from the EDE model, as demonstrated by the subsequent parameter constraints. 
Therefore, we anticipate that the $S_8$ prior approximation is applicable to the mentioned 
models in this paper, at least at the level of marginalized one-dimensional and two-dimensional 
posterior probability distributions \cite{PhysRevD.106.043525}.

\subsection{Results}
In order to assess the consistency among different datasets, we first examine the fitting performance 
of various models on all datasets \textit{except} for SH0ES. The results are presented in Tab.~\ref{tab:1}. 
The upper section of the table displays the 
parameters used for the MCMC sampling, while the lower section presents the derived parameters.
\begin{table*}
    \centering
    \caption{
    By excluding the SH0ES data, the best-fit values and marginalized posterior probabilities 
    at a 68\% confidence level for parameter constraints of the $\Lambda$CDM model, EDE model, 
    MCDS model, and YCDS model are obtained, using only CMB, BAO, SNIa, and $S_8$ 
    measurements obtained from DES-Y3 data.}
    \label{tab:1}
    \renewcommand{\arraystretch}{1.2}
\begin{tabular} { l  c  c  c  c}
    \hline
    \hline
    Model  &  $\Lambda$CDM  &  EDE  &  MCDS&  YCDS\\
    \hline
    {\boldmath$100\omega{}_\mathrm{b }$}&
    $2.249(2.252\pm 0.014)$&
    $2.283(2.279\pm 0.018)$&
    $2.276(2.272\pm 0.016)$&
    $2.272(2.272\pm 0.015)$\\

    {\boldmath$\omega{}_\mathrm{c}$}&
    $0.11823(0.11825\pm 0.00085)$&
    $0.1232(0.1230^{+0.0020}_{-0.0023})$&
    $0.1228(0.1221^{+0.0017}_{-0.0025})$&
    $0.1232(0.1225^{+0.0018}_{-0.0024})$\\

    {\boldmath$H_0$}&
    $68.11(68.21\pm 0.39)$&
    $69.74(69.94^{+0.79}_{-0.49})$&
    $69.53(69.47\pm 0.71)$&
    $69.51(69.65\pm 0.61)$\\

    {\boldmath$\ln(10^{10}A_\mathrm{s})$}&
    $3.045(3.045\pm 0.017)$&
    $3.060(3.055^{+0.015}_{-0.019})$&
    $3.050(3.049\pm 0.016)$&
    $3.040(3.049\pm 0.015)$\\

    {\boldmath$n_\mathrm{s}$}&
    $0.9699(0.9693\pm 0.0038)$&
    $0.9845(0.9813^{+0.0056}_{-0.0048})$&
    $0.9794(0.9777\pm 0.0058)$&
    $0.9766(0.9787\pm 0.0047)$\\

    {\boldmath$\tau{}_\mathrm{reio}$}&
    $0.0568(0.0563\pm 0.0080)$&
    $0.0594(0.0577^{+0.0081}_{-0.010})$&
    $0.0579(0.0563\pm 0.0077)$&
    $0.0509(0.0555\pm 0.0079)$\\

    {\boldmath$\log_{10}(m_{\phi})$}&
    $-$&
    $-26.837(-26.90^{+0.10}_{-0.081})$&
    $-26.824(-26.89^{+0.15}_{-0.12})$&
    $-26.920(-26.92^{+0.12}_{-0.067})$\\

    {\boldmath$\log_{10}(f_{\phi})$}&
    $-$&
    $26.392(26.405^{+0.081}_{-0.056})$&
    $26.381(26.348\pm 0.079)$&
    $26.361(26.369^{+0.094}_{-0.11})$\\

    {\boldmath$\alpha_\mathrm{i} $}&
    $-$&
    $2.845(2.81^{+0.10}_{-0.076})$&
    $2.843(2.80^{+0.10}_{-0.075})$&
    $2.806(2.76^{+0.13}_{-0.10})$\\    

    {\boldmath$\beta/\xi$}&
    $-$&
    $-$&
    $0.010(-0.0095\pm 0.041)$&
    $-0.014(-0.011\pm 0.018)$\\ 
    
    % \hline

    $10^{-9}A_\mathrm{s}$&
    $2.101(2.102\pm 0.035)$&
    $2.132(2.122^{+0.032}_{-0.041})$&
    $2.112(2.109^{+0.031}_{-0.035})$&
    $2.091(2.110\pm 0.032)$\\

    $100\theta{}_\mathrm{s}$&
    $1.04181(1.04204^{+0.00027}_{-0.00030})$&
    $1.04182(1.04183^{+0.00026}_{-0.00041})$&
    $1.04170(1.04182^{+0.00028}_{-0.00033})$&
    $1.04201(1.04185\pm 0.00040)$\\

    $f_\mathrm{EDE} $&
    $-$&
    $0.0548(0.056\pm 0.018)$&
    $0.0508(0.044^{+0.016}_{-0.022})$&
    $0.0502(0.050^{+0.018}_{-0.021})$\\

    $\log_{10}(z_\mathrm{c})$&
    $-$&
    $3.810(3.780^{+0.050}_{-0.037})$&
    $3.820(3.784^{+0.082}_{-0.074})$&
    $3.761(3.767^{+0.066}_{-0.038})$\\    

    $\Omega{}_\mathrm{m}$&
    $0.3047(0.3040\pm 0.0050)$&
    $0.3016(0.2995^{+0.0058}_{-0.0047})$&
    $0.3024(0.3015\pm 0.0057)$&
    $0.3033(0.3007\pm 0.0048)$\\

    $\sigma_8$&
    $0.8055(0.8056\pm 0.0065)$&
    $0.8216(0.8178\pm 0.0079)$&
    $0.8174(0.8121\pm 0.0092)$&
    $0.8121(0.8157^{+0.0067}_{-0.0076})$\\    

    $S_{8}$&
    $0.8118(0.8110\pm 0.0093)$&
    $0.8237(0.817\pm 0.010)$&
    $0.8206(0.814\pm 0.011)$&
    $0.8165(0.817\pm 0.010)$\\

    \hline
    % $\Delta\chi^2_\mathrm{tot}$  &  $1909.48$  &  $1911.14$  &  $1911.09$\\      
    $\chi^2_\mathrm{tot}$  &  $3818.96$  &  $3822.28$  &  $3820.16$&  $3820.40$\\      
    \hline
    \hline
\end{tabular}
\end{table*}

We observed that the constraints from the two coupling models closely align with those of 
the EDE model, yielding a slightly larger value for $H_0$ compared to the $\Lambda$CDM 
model's results. However, all models exhibit clear inconsistencies with the SH0ES data, 
consistent with previous research on EDE \cite{PRD.102.043507}.

Subsequently, we incorporated the SH0ES data and re-constrained various models.
The parameter constraint results for the $\Lambda$CDM model, the EDE model, the MCDS model, and the YCDS 
model are presented in Tab.~\ref{tab:2}. We employed the complete dataset, including CMB, 
BAO, SNIa, SH0ES, and $S_8$ from DES-Y3 data. 
\begin{table*}
    \centering
    \caption{The best-fit parameters and 68\% confidence level marginalized constraints 
    for the $\Lambda$CDM model, EDE model, MCDS model, and YCDS model are presented. The comprehensive 
    dataset, including CMB, BAO, SNIa, SH0ES, and $S_8$ from DES-Y3, is utilized. The upper 
    section of the table shows the cosmological parameters employed for MCMC sampling, 
    while the lower section displays the derived parameters.}
    \label{tab:2}
    \renewcommand{\arraystretch}{1.2}
\begin{tabular} { l  c  c  c  c}
    \hline
    \hline
    Model  &  $\Lambda$CDM  &  EDE  &  MCDS&  YCDS\\
    \hline
    {\boldmath$100\omega{}_\mathrm{b }$}&
    $2.260(2.263\pm 0.014)$&
    $2.276(2.281^{+0.024}_{-0.020})$&
    $2.280(2.287\pm 0.020)$&
    $2.268(2.278\pm 0.021)$\\

    {\boldmath$\omega{}_\mathrm{c}$}&
    $0.11729(0.11725\pm 0.00084)$&
    $0.1310(0.1299\pm 0.0028)$&
    $0.1287(0.1290^{+0.0028}_{-0.0023})$&
    $0.1293(0.1289\pm 0.0022)$\\

    {\boldmath$H_0$}&
    $68.64(68.71^{+0.35}_{-0.41})$&
    $71.85(72.46\pm 0.86)$&
    $72.23(72.20^{+0.93}_{-0.80})$&
    $72.23(72.19^{+0.78}_{-0.70})$\\

    {\boldmath$\ln(10^{10}A_\mathrm{s})$}&
    $3.047(3.050\pm 0.015)$&
    $3.057(3.063^{+0.015}_{-0.017})$&
    $3.065(3.064\pm 0.015)$&
    $3.064(3.063\pm 0.016)$\\

    {\boldmath$n_\mathrm{s}$}&
    $0.9733(0.9722\pm 0.0040)$&
    $0.9877(0.9908\pm 0.0059)$&
    $0.9898(0.9906^{+0.0057}_{-0.0051})$&
    $0.9849(0.9891^{+0.0053}_{-0.0048})$\\

    {\boldmath$\tau{}_\mathrm{reio}$}&
    $0.0576(0.0592\pm 0.0082)$&
    $0.0539(0.0563\pm 0.0090)$&
    $0.0565(0.0562\pm 0.0074)$&
    $0.0572(0.0571^{+0.0073}_{-0.0086})$\\

    {\boldmath$\log_{10}(m_{\phi})$}&
    $-$&
    $-27.292(-27.290\pm 0.055)$&
    $-27.292(-27.286^{+0.049}_{-0.60})$&
    $-27.333(-27.293^{+0.051}_{-0.062})$\\

    {\boldmath$\log_{10}(f_{\phi})$}&
    $-$&
    $26.632(26.616^{+0.056}_{-0.033})$&
    $26.609(26.602^{+0.047}_{-0.034})$&
    $26.630(26.613\pm 0.034)$\\

    {\boldmath$\alpha_\mathrm{i} $}&
    $-$&
    $2.762(2.783\pm 0.069)$&
    $2.772(2.774^{+0.067}_{-0.051})$&
    $2.722(2.738^{+0.076}_{-0.054})$\\    

    {\boldmath$\beta/\xi$}&
    $-$&
    $-$&
    $-0.001(-0.006\pm 0.014)$&
    $-0.012(-0.009\pm 0.016)$\\
    
    % \hline

    $10^{-9}A_\mathrm{s}$&
    $2.105(2.112\pm 0.032)$&
    $2.127(2.139^{+0.031}_{-0.036})$&
    $2.142(2.141\pm 0.031)$&
    $2.141(2.140^{+0.032}_{-0.036})$\\

    $100\theta{}_\mathrm{s}$&
    $1.04206(1.04217^{+0.00025}_{-0.00031})$&
    $1.04121(1.04145\pm 0.00043)$&
    $1.04172(1.04150\pm 0.00038)$&
    $1.04126(1.04149\pm 0.00035)$\\

    $f_\mathrm{EDE} $&
    $-$&
    $0.1183(0.119^{+0.023}_{-0.018})$&
    $0.1100(0.111^{+0.022}_{-0.014})$&
    $0.1181(0.114\pm 0.018)$\\

    $\log_{10}(z_\mathrm{c})$&
    $-$&
    $3.571(3.568\pm 0.034)$&
    $3.570(3.571\pm 0.029)$&
    $3.548(3.568^{+0.029}_{-0.035})$\\    

    $\Omega{}_\mathrm{m}$&
    $0.2983(0.2977\pm 0.0048)$&
    $0.2991(0.2923\pm 0.0056)$&
    $0.2915(0.2927^{+0.0061}_{-0.0051})$&
    $0.2925(0.2924\pm 0.0052)$\\

    $\sigma_8$&
    $0.8039(0.8047\pm 0.0060)$&
    $0.8329(0.8325\pm 0.0083)$&
    $0.8310(0.8294^{+0.0085}_{-0.0072})$&
    $0.8281(0.8305\pm 0.0078)$\\    

    $S_{8}$&
    $0.8016(0.8016^{+0.0096}_{-0.0080})$&
    $0.8316(0.822^{+0.011}_{-0.0093})$&
    $0.8192(0.819^{+0.011}_{-0.0087})$&
    $0.8177(0.820^{+0.012}_{-0.0087})$\\

    \hline
    $\chi^2_\mathrm{tot}$  &  $3838.20$  &  $3826.46$  &  $3825.94$  &  $3823.86$\\      
    % $\Delta\chi^2_\mathrm{tot}$  &  $-$  &  $-11.74$  &  $-12.26$  &  $-14.34$\\      
    $\Delta \mathrm{AIC}$        &  $-$  &  $ -5.74$  &  $ -4.26$  &  $ -6.34$\\ 
    \hline
    \hline
\end{tabular}
\end{table*}

We constrain the coupling constant $\beta$ ($\xi$) to be $-0.006\pm0.014$ ($-0.009\pm 0.016$) at a 68\% confidence 
level. This indicates a weak interaction between EDE and cold dark matter. 
The negative coupling constants align with our expectations, effectively increasing $H_0$ and reducing the 
matter power spectrum on small scales, as discussed in Sec.~\ref{sec:nr}.

The constrained values of $H_0$ for the MCDS model and the YCDS model are $72.20^{+0.93}_{-0.80}$ km/s/Mpc 
and $72.19^{+0.78}_{-0.70}$ km/s/Mpc, respectively, at a 68\% confidence level, both exceeding the value of $68.71^{+0.35}_{-0.41}$ km/s/Mpc 
for the $\Lambda$CDM model. This indicates that our coupling models inherit the ability of the EDE model to alleviate the Hubble tension. 

However, the $S_8$ values constrained by the MCDS model and YCDS model are 0.8192 and 
0.8177, respectively, which exacerbate the large-scale structure tension compared to 
the $\Lambda$CDM model's result of 0.8016. Nevertheless, the two coupling models have 
smaller $S_8$ values than the EDE model's result of 0.8316, partially mitigating the adverse effect caused 
by EDE.

\begin{figure}
	\includegraphics[width=\columnwidth]{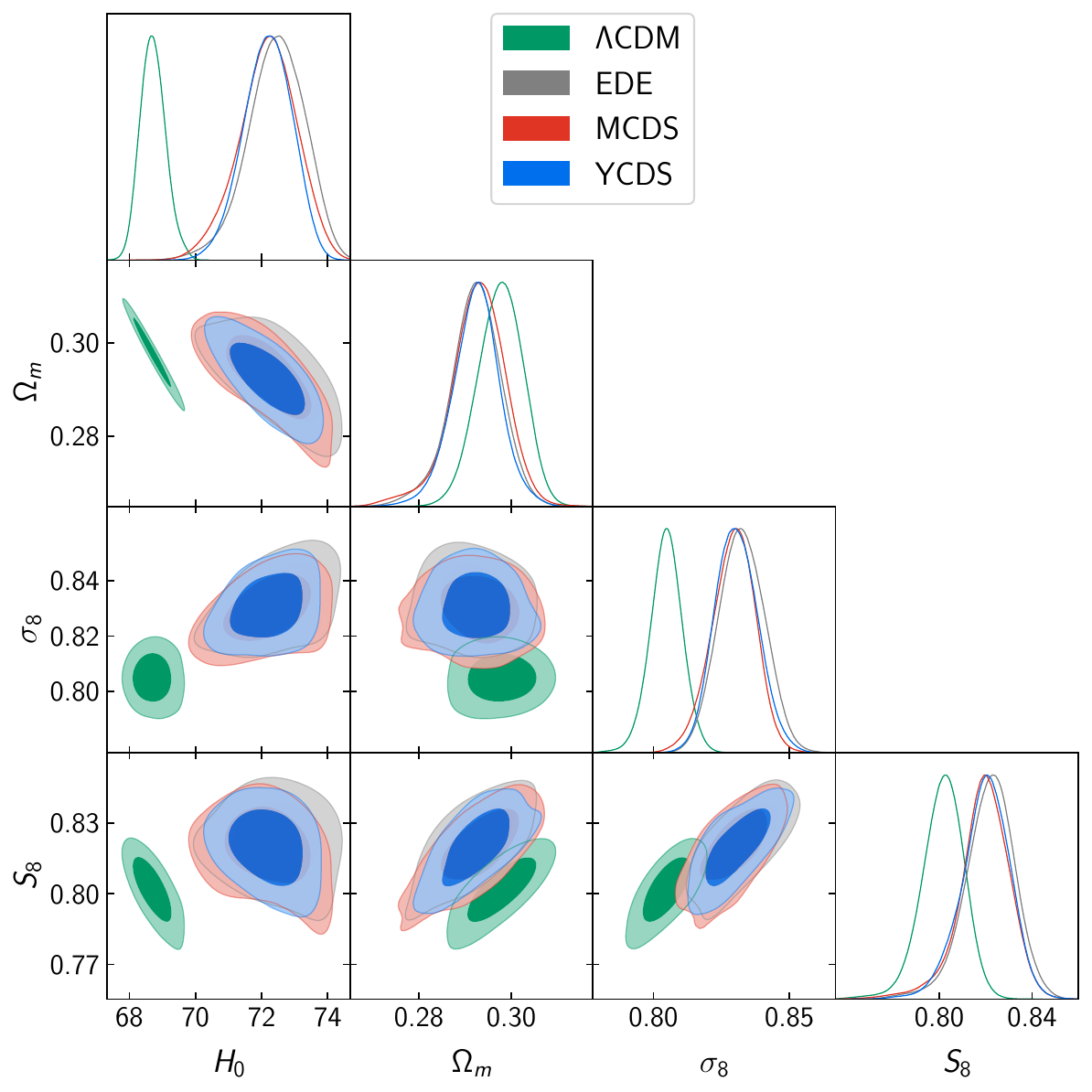}
    \caption{Posterior distribution plot for selected parameters in the four models are presented. 
        The MCDS model and YCDS model exhibit larger values of $H_0$ and $S_8$ relative to the 
        $\Lambda$CDM model, thereby alleviating the Hubble tension while exacerbating the 
        large-scale structure tension. However, the $S_8$ values of the two coupling models are smaller 
        than that of the EDE model, partially mitigating the negative effect of the EDE model.}
    \label{fig:7}
\end{figure}

Figure~\ref{fig:7} illustrates the posterior distribution plot for selected parameters in 
the four models (for complete posterior distributions, please refer to Fig.~\ref{fig:9} 
in the Appendix), revealing the noticeable increase in both $H_0$ and $S_8$ for the MCDS 
and YCDS models relative to the $\Lambda$CDM model. 

In addition, the interaction between dark matter and dark energy in the coupling models inhibit structure 
growth, thereby reducing the clustering effects of matter. Consequently, the MCDS model and 
the YCDS model result in smaller values of $S_8$ compared to the EDE model.

\begin{figure}
	\includegraphics[width=\columnwidth]{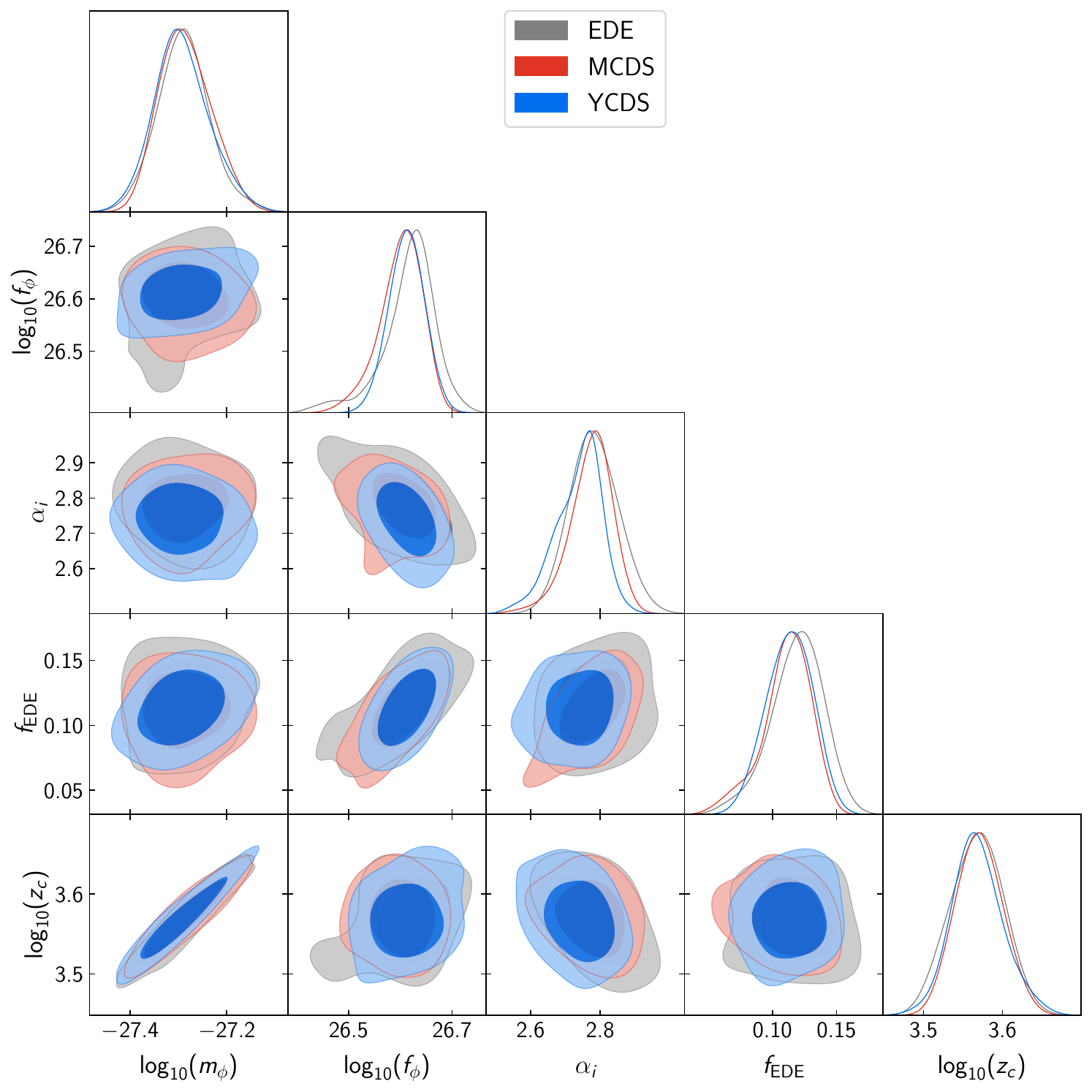}
    \caption{The posterior distributions of the EDE parameters for the EDE model and the 
    two coupling models are shown. The results from these three models are remarkably 
    consistent, with only minor deviations observed in the results of the coupling models 
    compared to those of the EDE model.}
    \label{fig:8}
\end{figure}

In Fig.~\ref{fig:8}, we present the posterior distributions of the EDE parameters for the EDE 
model and the two coupling models. We find that the results from these three models are remarkably 
consistent. In fact, the results for other cosmological parameters in the coupling models are also 
close to those of the EDE model, with only minor deviations.

The penultimate row of Tab.~\ref{tab:2} displays the $\chi^2_\mathrm{tot}$ values
of different models. It can be observed that the 
$\chi^2_\mathrm{tot}$ values for the EDE model, MCDS model, and YCDS model are both smaller than that 
of the $\Lambda$CDM model, with the $\Delta\chi^2_\mathrm{tot}$ values of $-11.74$, $-12.26$, and $-14.34$
respectively, primarily driven by the SH0ES data. The $\chi^2_\mathrm{tot}$ value of the 
MCDS model and YCDS model are smaller than that of the EDE model, owing to the $S_8$ data from DES-Y3. 

We also calculated the Akaike information criterion (AIC) to compare the models 
\cite{Akaike_1974} \footnote{In fact, the utilization of Bayesian evidence for model 
selection is preferable. AIC often tends to favor overly complex models, particularly in 
cases of small sample sizes or high noise levels. However, due to the complexity involved 
in computing Bayesian evidence, we opt to employ the AIC in this study.}, 
\begin{equation}
    \mathrm{AIC}=\chi^2_\mathrm{tot}+2k,
\end{equation}
where $k$ represents the number of fitting parameters. The AIC values for the EDE model, MCDS model, 
and YCDS model relative to the $\Lambda$CDM model are displayed in the final row of 
Tab.~\ref{tab:2}, which are $-5.74$, $-4.26$, and $-6.34$, respectively. Although the $\chi^2_\mathrm{tot}$ 
value of the MCDS model is smaller than that of the EDE model, the introduction of a new 
parameter results in a higher AIC value. The AIC value for the YCDS 
model is the smallest, indicating that, from the perspective of AIC, the YCDS model performs the best.

To quantify the level of tension using the SH0ES data, we calculated the following tension 
metric (in units of Gaussian $\sigma$) \cite{PhysRevD.99.043506,SCHONEBERG20221}, 
\begin{equation}
    Q_\mathrm{DMAP}\equiv \sqrt{\chi^2(w/\,\mathrm{SH0ES})-\chi^2(w/o\, \mathrm{SH0ES})},
\end{equation}
which involves the disparity in $\chi^2$ when considering the data with and without SH0ES.
This metric effectively captures the non-Gaussian nature of the posterior distribution. 
The tension metric yields results of 4.4$\sigma$, 2.1$\sigma$, 2.4$\sigma$, and 1.9$\sigma$ for 
the $\Lambda$CDM model, EDE model, MCDS model, and YCDS model, respectively. 
Based on this criterion, we consider the performance of the EDE model and the two coupling 
models to be superior to that of the $\Lambda$CDM model, with the YCDS model exhibiting 
the best performance.

\section{Conclusions}
\label{sec:con}
In this paper, we consider the interaction between early dark energy (EDE) and 
cold dark matter, proposing the momentum-coupled dark sector (MCDS) model and the 
Yukawa-coupled dark sector (YCDS) model to alleviate the 
Hubble tension and large-scale structure tension. The EDE component in the two coupling models is 
employed to alleviate the Hubble tension, while the momentum (or energy and momentum) exchange 
between EDE and cold dark matter can affect the evolution of cold dark matter density 
perturbation, thereby suppressing structure growth and mitigating large-scale structure 
tension.

We investigate the evolution equations of the background and perturbation for the coupled 
models, along with providing the corresponding initial conditions. We discuss the 
modifications to the original EDE model due to the momentum and Yukawa couplings between EDE and cold 
dark matter, as well as its effect on structure growth and matter power spectrum. 
Subsequently, we utilize various cosmological data, including CMB, BAO, SNIa, SH0ES, and 
$S_8$ from DES-3, to constrain the $\Lambda$CDM model, EDE model, MCDS model, and YCDS model.

We obtain the coupling constant $\beta$ ($\xi$) to be $-0.006\pm0.014$ ($-0.009\pm 0.016$) at a 68\% confidence level, the 
negative coupling constants can suppress structure growth on small scales, aiding in 
alleviating the large-scale structure tension. The values for $H_0$ in the 
MCDS model and YCDS model are $72.20^{+0.93}_{-0.80}$ km/s/Mpc 
and $72.19^{+0.78}_{-0.70}$ km/s/Mpc, respectively, at a 68\% confidence level, 
both models can alleviate the Hubble tension. 

Meanwhile, the constrained values of $S_8$ in the two coupling models are 0.8192 and 0.8177, 
respectively, exceeding the results of the $\Lambda$CDM model, further exacerbating the 
large-scale structure tension. However, the interaction between EDE and cold dark matter 
in the MCDS model and the YCDS model lead to smaller values of $S_8$ compared to the EDE model's result of 0.8316, partially 
mitigating the negative effect of the original EDE model.

We compared the $\chi^2_\mathrm{tot}$ values of different models, where the 
$\chi^2_\mathrm{tot}$ values for the EDE model, MCDS model, and YCDS model relative to the $\Lambda$CDM 
model are $-11.74$, $-12.26$, and $-14.34$, respectively. The YCDS model exhibited the lowest 
$\chi^2_\mathrm{tot}$ value. Additionally, we 
calculated the Akaike information criterion (AIC) for model comparison, with the results 
being $-5.74$, $-4.26$, and $-6.34$ for the EDE model, MCDS model, 
and YCDS model relative to the $\Lambda$CDM model, respectively. The YCDS model exhibits 
the smallest AIC value, indicating that, based on the AIC, it delivers the 
best performance among the models considered.

The two coupling models preserve the partially mitigation of the Hubble tension achieved by the 
EDE model, but they still fall short of completely resolving the large-scale structure 
tension. However, the couplings between EDE and cold dark matter alleviate the negative 
effect of the original EDE model, resulting in a smaller $S_8$ compared to the EDE model. 
Further research is needed to fully address the cosmological tensions.

\begin{acknowledgments}
    This work is supported in part by National Natural Science Foundation of China 
    under Grant No.12075042, Grant No.11675032 (People's Republic of China).
\end{acknowledgments}

\appendix*
\section{The full MCMC posteriors}
\begin{figure*}
	\includegraphics[width=\linewidth]{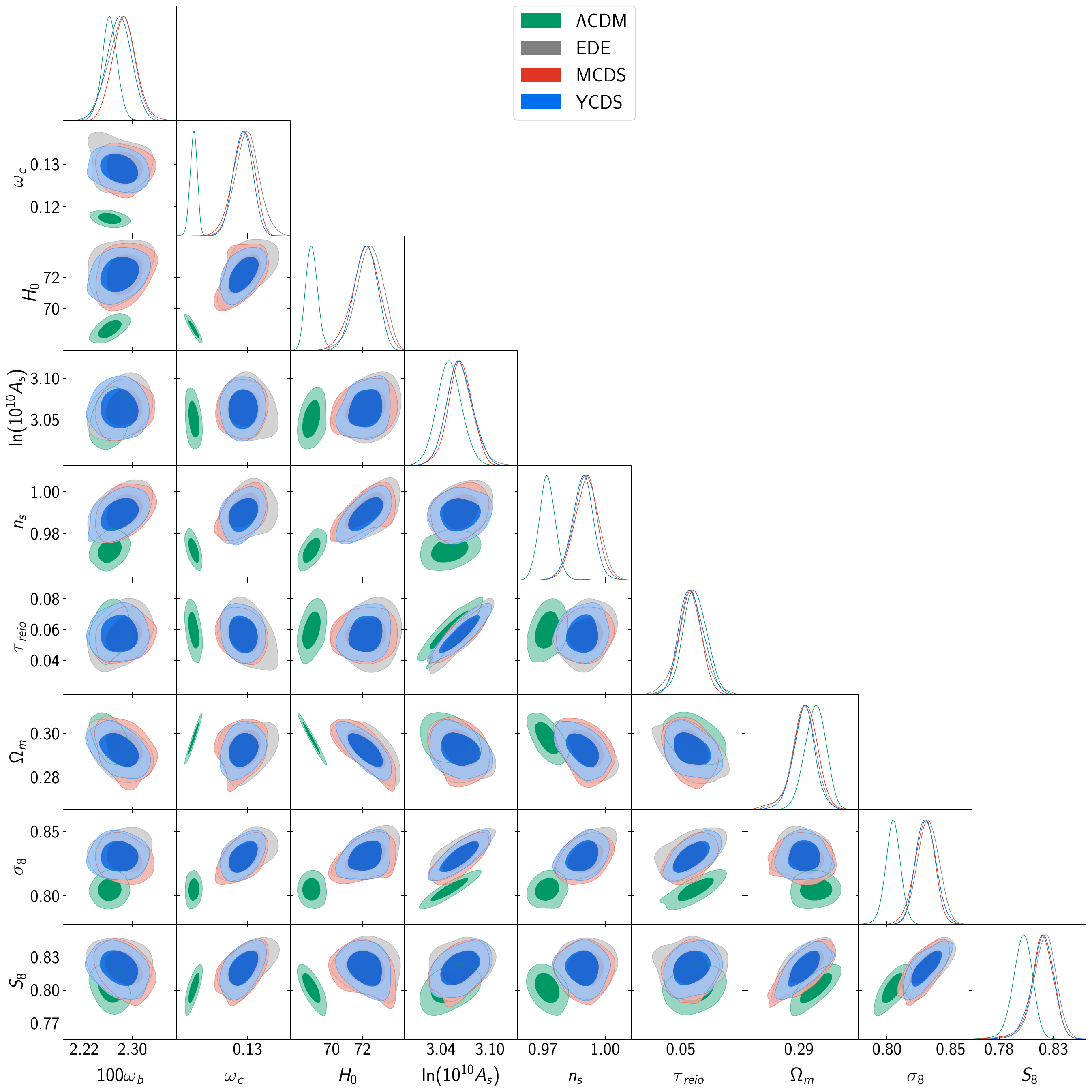}
    \caption{The comprehensive posterior distributions for the $\Lambda$CDM, EDE, MCDS, and YCDS 
        models are provided, utilizing data encompassing CMB, BAO, SNIa, SH0ES, and 
        $S_8$ from DES-Y3.}
    \label{fig:9}
\end{figure*}

\bibliography{cds}% Produces the bibliography via BibTeX. 

\end{document}